\documentclass[%
 reprint,
 amsmath,amssymb,
 aps,
]{revtex4-1}
\usepackage{caption}
\usepackage{graphicx}
\usepackage{epstopdf}
\usepackage{dcolumn}
\usepackage{bm}
\usepackage{balance}

\begin{document}


\title{Resonant photoproduction of high-energy electron-positron pairs in the field of a nucleus and a plane electromagnetic wave }

\author{Nikita R. Larin}
 \email{nikita.larin.spb@gmail.com}
\affiliation{
 Department of Theoretical Physics Peter the Great St. Petersburg Polytechnic University\\
St-Petersburg, Russia}

\author{Victor V. Dubov}
 \email{dubov@spbstu.ru}
 \affiliation{
 Department of Theoretical Physics Peter the Great St. Petersburg Polytechnic University\\
 St-Petersburg, Russia}

\author{Sergei P. Roshchupkin}
 \email{serg9rsp@gmail.com}
\affiliation{
 Department of Theoretical Physics Peter the Great St. Petersburg Polytechnic University\\
 St-Petersburg, Russia}

\date{\today}

\begin{abstract}
The resonant photoproduction of ultrarelativistic electron-positron pairs (PPP) in a nuclear field and a weak laser field is theoretically studied. Under resonance conditions, the intermediate virtual electron (positron) in the laser field becomes a real particle. As a result, the initial process of the second order in the fine structure constant in the laser field effectively reduces into two successive processes of the first order: single-photon production of electron-positron pair in a laser field (laser-stimulated Breit-Wheeler process) and laser-assisted process of electron (positron) scattering on a nucleus. Resonant kinematics of PPP is studied in details. It is shown that for the considered laser intensities resonance is possible only for the initial photon energies greater than the characteristic threshold energy. At the same time, the ultrarelativistic electron and positron propagate in a narrow cone along the direction of the initial photon momentum. The resonant energy of the positron (electron) can has two values for each radiation angle which varies from zero to some maximum value determined by the energy of the initial photon and the threshold energy. Resonant differential cross section of the studied process was obtained. It is shown that the resonant differential cross section of the PPP can significantly exceed the corresponding cross section of the PPP without an external field. The project calculations may be experimentally verified by the scientific facilities of pulsed laser radiation (SLAC, FAIR, XFEL, ELI, XCELS).
\end{abstract}

\maketitle


\section{\label{sec:level1}Introduction}
Due to the use of powerful laser sources in modern applied and fundamental research [1-5] a theoretical study of quantum electrodynamics (QED) processes in strong light fields seems to be one of the  priorities, which is developing intensively (see, for example, [6-46]). The main research results were systematized in monographs. [11-14] and reviews [15-21]. 

It is important to emphasize that QED processes of higher orders in the fine structure constant in a laser field (laser-assisted QED processes) can occur both non-resonant and resonant channels. In laser field can take place so called Oleinik's resonances [9,10] associated with the fact that in the light field  lower order processes in the fine structure constant are allowed (laser-stimulated QED processes) [15]. It is important to note that the probability of the resonant QED processes in a laser field can significantly (by several orders of magnitude) exceed the corresponding probability of a process without an external field.

Underline that the Oleinik's resonances for the PPP on the nucleus in the wave field were studied only for one of the possible channels, when the initial gamma-quantum decay into a positron and an intermediate electron in the wave field, which is then scattered by the nucleus. The second channel, when the initial photon decay into an electron and an intermediate positron, which is then scattered at the nucleus has not been studied. We also emphasize that for the first channel, considered the case of an ultrarelativistic pair, when the positron propagates in a narrow cone with the direction of the initial photon momentum, and the electron is scattered at large angles [12-14, 19-20].

In this paper, the theory of the resonant PPP of ultrarelativistic energies is developed for the case when a positron and an electron propagate in a narrow cone along the direction of the momentum of the initial gamma-quantum, i.e. with taking into account channels a) and b) (see Fig.2). The process is considered in the Born approximation for interaction with the field of the nucleus. ($v/c\gg Z/137$ , $v$ - electron velocity, $c$ - speed of light in vacuum,  $Z$ - charge of the nucleus).

Point out that there are two characteristic parameters in a problem of PPP. Classical relativistically invariant parameter [15, 19-22]:
\begin{equation} \label{eq:1}
\eta  = \frac{{eF\mathchar'26\mkern-10mu\lambda  }}{{m{c^2}}}
\end{equation}
numerically equal to the ratio of the work of the field at a wavelength to the rest energy of an electron  (where $e$ and $m$ - are the charge and mass of an electron, $F$ and $\mathchar'26\mkern-10mu\lambda = c/\omega$ - are strength of electric field and wavelength correspondently, $\omega$ - is the frequency of a wave) and quantum multiphoton parameter  (Bunkin-Fedorov parameter) [8, 11, 19-22]:
\begin{equation} \label{eq:2}
\gamma_i = \eta \frac {m v_i c}{\hbar \omega}
\end{equation}
Herein $v_i$  - is the velocity of the initial electron. Within the optical range of the frequencies ($\omega \sim 10^{15} s^{-1}$) the classical parameter is $\eta \sim 1$ for the fields of $F \sim 10^{10} \div 10^{11} V/cm$, quantum multiphoton parameter is $\gamma_i \sim 1$ for the fields of $F \sim (10^5 \div 10^6)(c/v_i) V/cm$. Therefore, when $\eta \ll 1$ the quantum multiphoton parameter  $\gamma_i$ may be large. However, this is true only when the electrons (positrons) are scattered on the nucleus at large angles. In this case, the main parameter determining multiphoton processes is the Bunkin-Fedorov quantum parameter. Accordingly, the problem is usually studied in the intensity of moderately strong fields, in which these parameters satisfy the following conditions:
\begin{equation} \label{eq:3}
\eta \ll 1 , 
\end{equation}
and $\gamma_i \gtrsim 1$. It is important that for the process of PPP with the scattering of electrons (positrons) on the nucleus at small angles, the Bunkin-Fedorov quantum parameter does not appear [44]. In this case, the main parameter of multiphoton processes is the classical parameter $\eta$ and in conditions (3) the laser field can be considered as weak [44]

Further in the article we use the relativistic system of units: $\hbar=c=1$.

\section{\label{sec:level1}The amplitude of the PPP on a nucleus in a light field}

Let us choose a 4-potential of an external elliptically polarized light wave propagating along the axis $z$  in the following form:
\begin{equation} \label{eq:4}
A({\phi})={\frac {F} {\omega}} {\cdot} (e_x \cos{\phi} + \delta e_y \sin{\phi}), \: \phi=kx={\omega(t-z)}
\end{equation}
Here $\delta$ - is the ellipticity parameter ($\delta = 0$ - the linear polarization, $\delta = \pm 1$ - the circular polarization), $e_{x,y}=(0,{\bf e}_{x,y})$ and $k={\omega n}={\omega (1,{\bf n})}$ - 4-vectors of polarization and the momentum of the electromagnetic wave, particularly: $k^2=0, \: e_{x,y}^2=-1, \: (e_{x,y}k)=0$.

We study the problem of PPP on the nucleus in the field of a plane electromagnetic wave in the Born approximation on the interaction of electrons and positrons with the field of the nucleus. This is a second order process in the fine structure constant and it is described by two Feynman diagrams. (see Fig.1).

\begin{figure}[h!]
     \begin{center}
     \includegraphics[width=8cm]{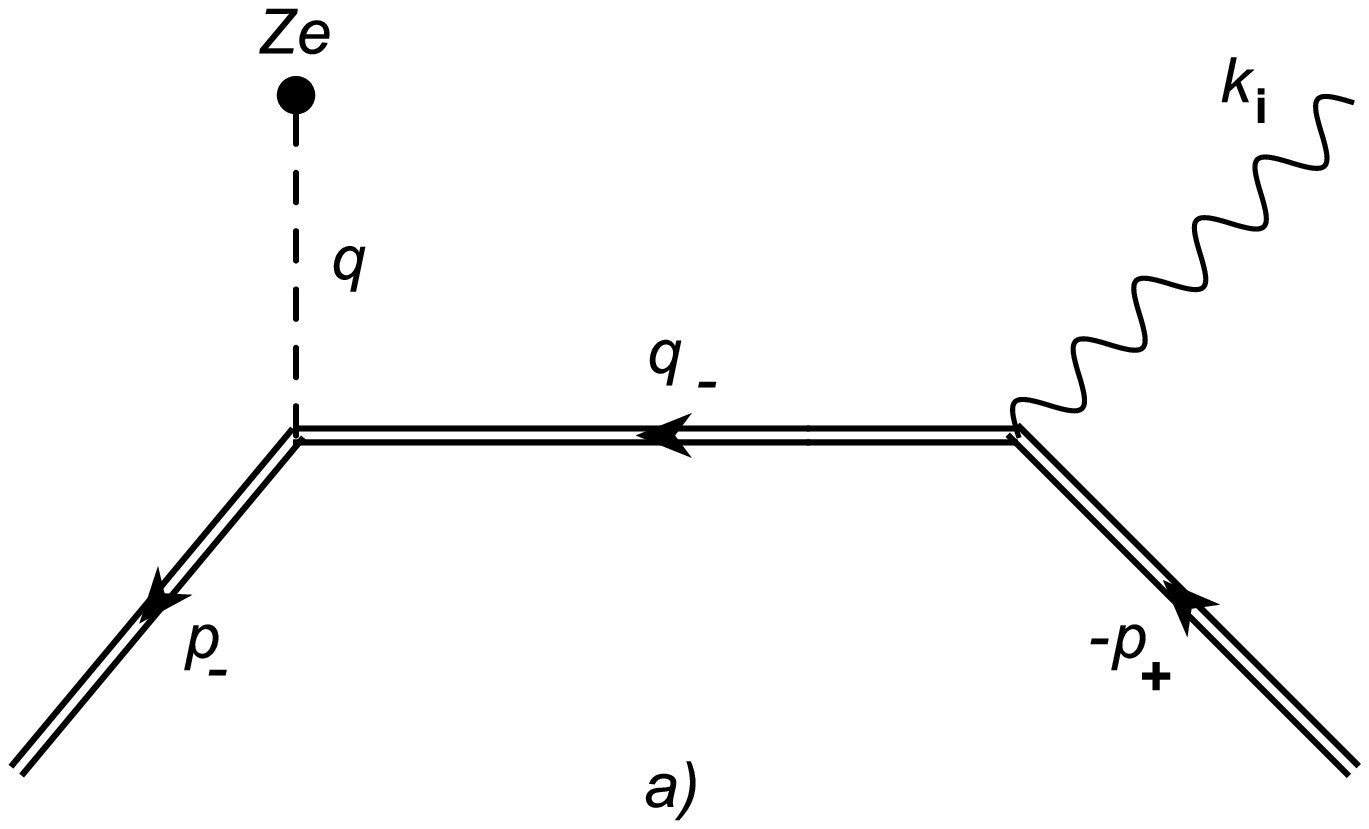}
     \includegraphics[width=8cm]{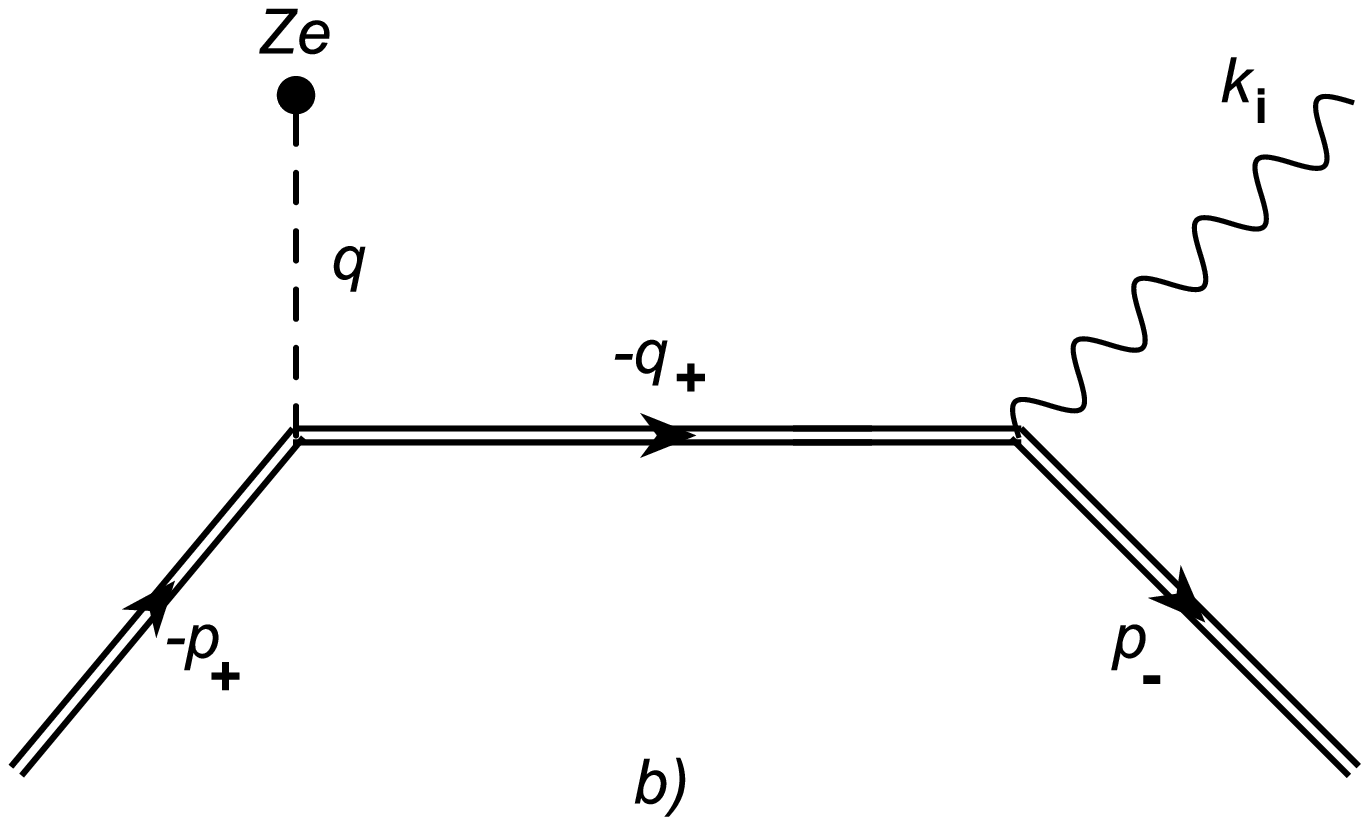}
     \end{center}
     \caption{Feynman diagrams of the PPP process on the nucleus in the field of a plane electromagnetic wave. Double incoming and outgoing lines correspond to the Volkov functions of the electron and positron in the initial and final states, the inner line corresponds to the Green function of the fermion in the field of a plane wave (4). Wavy line corresponds to 4-momentum of incident gamma quantum. Dashed line stands for 4-momentum of pseudo photon}
  \end{figure}

 The amplitude of such process after simple calculations can be represented as follows. (see [12-14, 18, 19, 27]):
\begin{equation} \label{eq:5}
S_{fi} = \sum_{l=-\infty}^{\infty} S_l
\end{equation}
where the partial amplitude with emission and absorption of  $|l|$ - photons of the wave has the following form:
\begin{equation} \label{eq:6}
{S_l} = i \cdot \frac{{8Z{e^3}{\pi ^{5/2}}}}{{\sqrt {2{E_ - }{E_ + }{\omega _i}} }} \cdot \exp (i\tilde \phi ) \cdot \left[ {\bar u{M_l}v} \right] \cdot \frac{{\delta ({q_0})}}{{{{\bf{q}}^2}}},
\end{equation}
\begin{equation} \label{eq:7}
\begin{aligned}
M_l = \sum_{l_1=-\infty}^{\infty} \left[H_{l+l_1}^0 ({{{\tilde p}_{ - }}}, q_-) \cdot {\frac {\hat{q}_- + m_{*}} {q^2_- - m^2_{*}}} \cdot  \varepsilon_\mu H_{l_1}^{\mu}(q_-,{{{\tilde p}_{ + }}})\right. + \\
\left.{ + {\varepsilon _\mu }H_{{l_1}}^\mu ({{\tilde p}_ - },{q_{ + ,}}) \cdot \frac{{{{\hat q}_ + } + {m_*}}}{{q_ + ^2 - m_*^2}} \cdot H_{l + {l_1}}^0({q_ + },{{\tilde p}_ + })}\right]
\end{aligned}
\end{equation}

In the relations (6)-(7) $\tilde \phi $ - is the independent from the summation indexes phase, $\varepsilon_\mu$ - is the 4-vector of polarization of incident photon $v,\overline{u}$ and ${\tilde p_ \pm } = \left( {{{\tilde E}_ \pm },{{{\bf{\tilde p}}}_ \pm }} \right)$ - are the Dirac bispinors and the 4-quasimomenta of the positron and electron. The 4-momenta of the intermediate positron and electron $q_{\pm}$ and the transmitted 4-momentum $q$ are determined by expressions :
\begin{equation} \label{eq:8}
{q_ - } = {k_i} + k - {l_1}{\tilde p_ + }, \; {q_ + } = {k_i} + k - {l_1}{\tilde p_ - }
\end{equation}
\begin{equation} \label{eq:9}
q = {\tilde p_ + } + {\tilde p_ - } - {k_i} - lk
\end{equation}
Here $k_i = \omega_i (1,{\bf n_i})$ - is the 4-momentum of the initial photon, $m_{*}$ - is the effective mass of an electron in the light field [15, 47, 48]:
\begin{equation} \label{eq:10}
\tilde{p}_{\pm} = p_{\pm} + \left(1 + \delta^2\right) \eta^2 {\frac {m^2} {4(k p_{\pm})}} k,
\end{equation}
\begin{equation} \label{eq:11}
\tilde{p}_{\pm}^2 = m^2_{*}, \: m_{*} = m {\sqrt{1 + {\frac {1} {2}}\left(1 + \delta^2\right) \eta^2}}
\end{equation}
Here ${p_ \pm } = \left( {{E_ \pm },{{\bf{p}}_ \pm }} \right)$ are 4-momenta of positron and electron. Notation with the hat in relation  (7) and further stands for the dot product of the corresponding 4-vector with Dirac gamma matrices: ${\tilde \gamma ^\mu } = \left( {{{\tilde \gamma }^0},{\bf{\tilde \gamma }}} \right)$, $\mu  = 0,1,2,3$. For example, ${\hat q_ + } = q_ + ^\mu {\tilde \gamma _\mu } = q_ + ^0{\tilde \gamma _0} - {{\bf{q}}_ + }{\bf{\tilde \gamma }}$. The amplitudes $H_{{l_1}}^\mu $ and $H_{l + {l_1}}^0$ (see Fig.1) in relation (7) have the following form:
\begin{equation} \label{eq:12}
\begin{aligned}
H_{{l_1}}^\mu \left( {{p_2},{p_1}} \right) = {a^\mu }{L_{{l_1}}}\left( {{p_2},{p_1}} \right) + b_ - ^\mu {L_{{l_1} - 1}}+ \\  + b_ + ^\mu {L_{{l_1} + 1}} + {c^\mu }({L_{{l_1} + 2}} + {L_{{l_1} - 2}})
\end{aligned}
\end{equation}
\begin{equation} \label{eq:13}
\begin{aligned}
H_{l + {l_1}}^0\left( {{p_2},{p_1}} \right) = {a^0}{L_{l + {l_1}}}\left( {{p_2},{p_1}} \right) + b_ - ^0{L_{l + {l_1} - 1}}+ \\  + b_ + ^0{L_{l + {l_1} + 1}} + {c^0}({L_{l + {l_1} + 2}} + {L_{l + {l_1} - 2}})
\end{aligned}
\end{equation}

matrices ${a^\mu },b_ \pm ^\mu ,{c^\mu }$ are determined by the expressions:
\begin{equation} \label{eq:14}
a^{\mu} = \tilde{\gamma}^{\mu} + \left(1+{\delta}^2\right) \cdot {\eta}^2 {\frac {m^2} {4(kp_1)(kp_2)}} k^{\mu}
\end{equation}
\begin{equation} \label{eq:15}
b^{\mu}_{\pm} = {\frac {1} {4}} \eta m \cdot \left[{\frac {\hat{\varepsilon}_{\pm} \hat{k} \tilde{\gamma}^{\mu}} {(kp_2)}} + {\frac {\tilde{\gamma}^{\mu} \hat{k}  \hat{\varepsilon}_{\pm} } {(kp_1)}}\right], \: \hat{\varepsilon}_{\pm} = \hat{e}_{x} \pm i \delta \cdot \hat{e}_{y}
\end{equation}
\begin{equation} \label{eq:16}
c^{\mu} = -\left (1 -{\delta}^2\right) \cdot {\eta}^2 {\frac {m^2} {8(kp_1)(kp_2)}} \cdot k^{\mu}
\end{equation}
Special functions ${L_l}$  and their arguments have the following form [46]:
\begin{equation} \label{eq:17}
 {L_l}\left( {\gamma ,\beta ,\chi } \right) = e^{-il\chi}\sum\limits_{n' =  - \infty }^\infty  {e^{2in'\chi}{J_{l - 2n'}}\left( \gamma  \right)} {J_{n'}}\left( \beta  \right),
\end{equation}
\begin{equation} \label{eq:18}
tan \chi = \delta \cdot {\frac {\left(e_y Q\right)} {\left(e_x Q\right)}}, \: Q = {\frac {p_2} {\left(kp_2\right)}} - {\frac {p_1} {\left(kp_1\right)}}
\end{equation}
\begin{equation} \label{eq:19}
\gamma = \eta m \sqrt{(e_x Q)^2 + \delta^2 \cdot (e_y Q)^2}
\end{equation}
\begin{equation} \label{eq:20}
\beta = {\frac {1} {8}} (1-\delta^2) \eta^2 m^2 \cdot \left[{\frac {1} {(kp_2)}} - {\frac {1} {(kp_1)}}\right]
\end{equation}
For the amplitudes $H_{l + {l_1}}^0({\tilde p_{ - ,}}{q_ - })$ and $H_{l + {l_1}}^0({q_ + },{\tilde p_ + })$ in expressions (13), (14) -(20) we have to assume ${p_1} \to {q_ - },\;{p_2} \to {\tilde p_ - }$ and ${p_1} \to {\tilde p_ + },\:{p_2} \to {q_ + }$ and for amplitudes $H_{{l_1}}^\mu ({q_ - },{\tilde p_{ + ,}})$ and $H_{{l_1}}^\mu ({\tilde p_ - },{q_{ + ,}})$ in relations (12), (14)-(20)  have to make a transform ${p_1} \to {\tilde p_ + },\:{p_2} \to {q_ - }$ and ${p_1} \to {q_ + },\:{p_2} \to {\tilde p_ - }$.

\section{\label{sec:level1}The poles of the PPP amplitude}

Resonant behavior of the studied process explained by the fact that low-order processes in the fine structure constant in the field of the electromagnetic wave are allowed, since intermediate particle enters the mass shell [12-14, 18, 19, 27]  (see Fig.2). 
\begin{equation} \label{eq:21}
q_ - ^2 = {m^2},\quad q_ + ^2 = {m^2}
\end{equation}
where 4-momenta of intermediate electron $q_-$  and positron $q_+$ under conditions (3)  define as follows:
\begin{equation} \label{eq:22}
{q_ - } = {k_i} + k - {p_ + },{\rm{  }}{q_ + } = {k_i} + k - {p_ - }
\end{equation}
\begin{figure}[h!]
     \begin{center}
     \includegraphics[width=8cm]{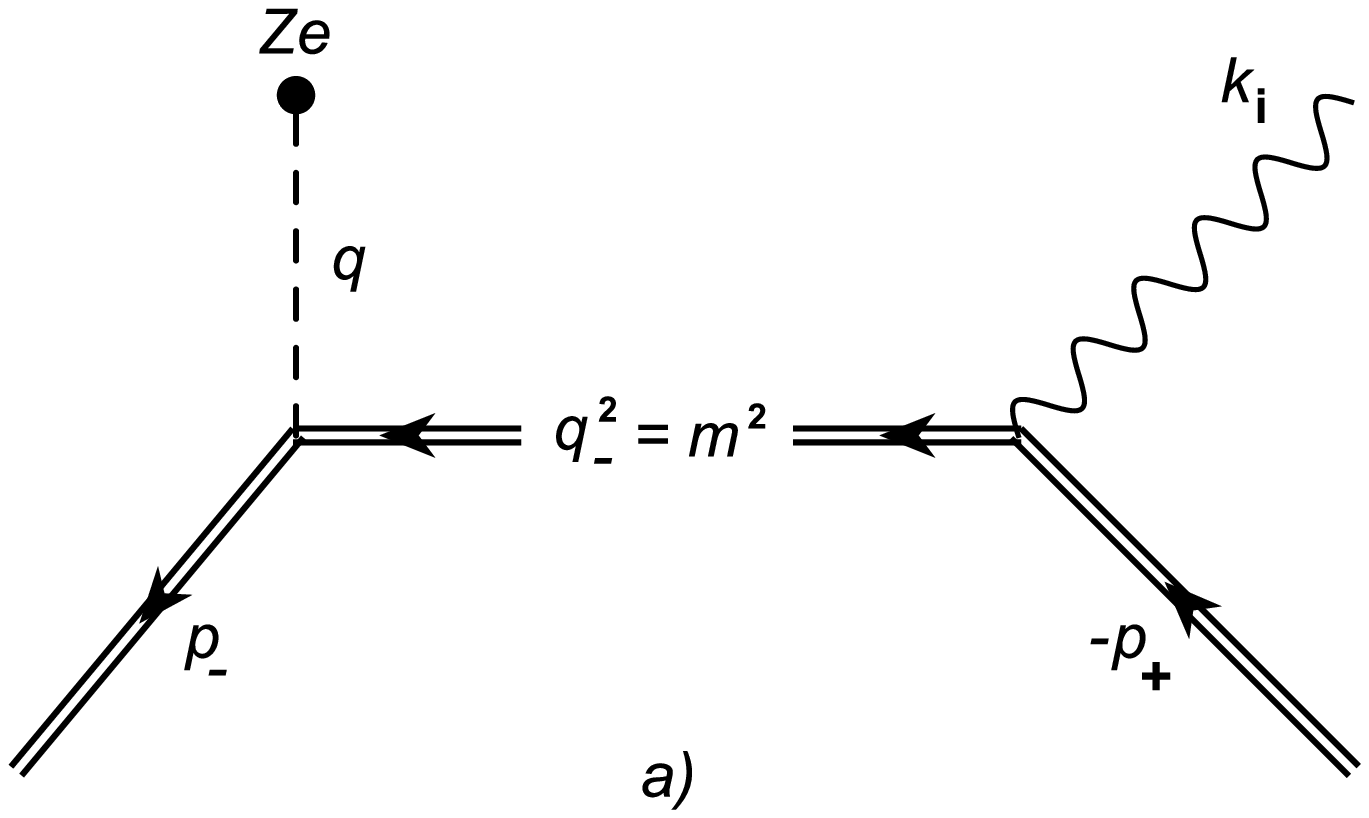}
     \includegraphics[width=8cm]{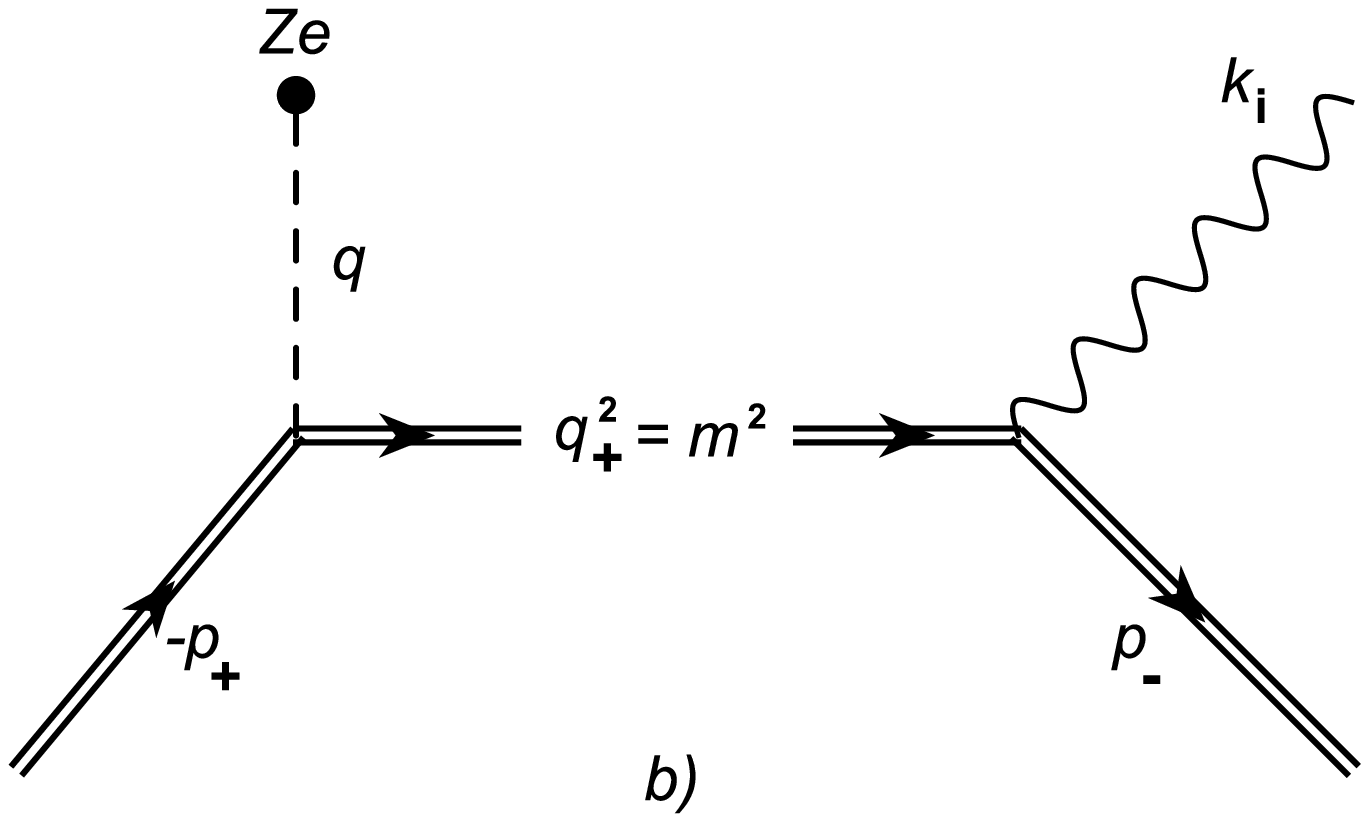}
     \end{center}
     \caption{Resonant photoproduction of electron-positron pair in the field of the nucleus and a plane electromagnetic wave.}
  \end{figure}

It is easy to verify that with the coincidence of the directions of propagation of the initial photon and external field the simultaneous fulfillment of the resonant conditions (21) and (22) is impossible. We also note that for the process of electron-positron pair production by a photon in the field of a plane electromagnetic wave the main parameter is the classical relativistically invariant parameter $\eta$. Therefore for the fields (3) the most probable processes with absorption of one photon from the external field ${l_1} = 1$. At the same time, for the process of electron or positron scattering on the nucleus at large angles in the field of a wave, the main parameter is the Bunkin-Fedorov quantum parameter $\gamma_i\gtrsim1$. Because of this, electron or positron can emit and absorb a large number of photons of the external field. However, if electron or positron is scattered at small angles, then the classical parameter   becomes the main parameter for the process of electron or positron scattering on the nucleus and under conditions (3) the process with one photon of the laser field is most probable. 

Considering expressions (21) and (22) it is not difficult to obtain relations for the initial photon frequency at resonance for diagrams a) and b) (see Fig.2). Under conditions (3) we have:
\begin{equation} \label{eq:23}
{\omega _i}_{\left( a \right)} = {l_1}\omega \frac{{{\kappa _ + }}}{{{\kappa _{i + }} + 2{l_1}\omega {{\sin }^2}\left( {{{{\theta _i}} \mathord{\left/
 {\vphantom {{{\theta _i}} 2}} \right.
 \kern-\nulldelimiterspace} 2}} \right)}}
\end{equation}
\begin{equation} \label{eq:24}
{\omega _i}_{\left( b \right)} = {l_1}\omega \frac{{{\kappa _ - }}}{{{\kappa _{i - }} + 2{l_1}\omega {{\sin }^2}\left( {{{{\theta _i}} \mathord{\left/
 {\vphantom {{{\theta _i}} 2}} \right.
 \kern-\nulldelimiterspace} 2}} \right)}}
\end{equation}
Herein
\begin{equation} \label{eq:25}
{\kappa _{i \pm }} = {E_ \pm } - \left| {{{\bf{p}}_ \pm }} \right|\cos {\theta _{i \pm }},\; {\kappa _ \pm } = {E_ \pm } - \left| {{{\bf{p}}_ \pm }} \right|\cos {\theta _ \pm },
\end{equation}
\begin{equation} \label{eq:26}
{\theta _ \pm } = \left( {{\bf{k}},{{\bf{p}}_ \pm }} \right),\quad {\theta _{i \pm }} = \left( {{{\bf{k}}_i},{{\bf{p}}_ \pm }} \right),\quad {\theta _i} = \left( {{{\bf{k}}_i},{\bf{k}}} \right)
\end{equation}

Further, we study the case of high energies of the initial gamma-quantum, when the electron-positron pair is ultrarelativistic and propagates in a narrow cone along the momentum of the initial photon. 
\begin{equation} \label{eq:27}
{E_ \pm } >  > m,
\end{equation}
\begin{equation} \label{eq:28}
{\theta _{i \pm }} \ll 1,\quad \theta  = \left( {{{\bf{p}}_ + },{{\bf{p}}_ - }} \right) \ll1,\quad {\theta _i}\sim1,
\end{equation}
Under conditions (27), (28), and also by virtue of the law of conservation of energy in the external field (3) ${\omega _i} \approx {E_ + } + {E_ - }$ we can get following relations:
\begin{equation} \label{eq:29}
{\kappa _ \pm } \approx 2{E_ \pm }{\sin ^2}\frac{{{{\theta }_ \pm }}}{2},\quad {\kappa _{i \pm }} \approx \frac{{{m^2}}}{{2{E_ \pm }}}\left( {1 + \tilde \delta _ \pm ^2} \right),
\end{equation}
Denoted here
\begin{equation} \label{eq:30}
\begin{array}{l}
{{\tilde \delta }_ \pm } = \frac{{{E_ \pm }{\theta _{i \pm }}}}{m} = 2{x_ \pm }{\delta _ \pm },\; {x_ \pm } = \frac{{{E_ \pm }}}{{{\omega _i}}},\; {\delta _ \pm } = \frac{{{\omega _i}{\theta _{i \pm }}}}{{2m}}.
\end{array}
\end{equation}

Taking into account (25) - (30), and (23),(24) it is easy to obtain possible values of the resonant energies of the electron and positron for channels a) and b):
\begin{equation} \label{eq:31}
\begin{array}{l}
{x_{\left( a \right) + }}\left( {\delta _ + ^2} \right) = \frac{1}{{2\left( {\delta _ + ^2 + {\varepsilon _i}} \right)}}\cdot\left[ {{\varepsilon _i} \pm \sqrt {{\varepsilon _i}\left( {{\varepsilon _i} - 1} \right) - \delta _ + ^2} } \right],
\end{array}
\end{equation}
\begin{equation} \label{eq:32}
\begin{array}{l}
{x_{\left( b \right) - }}\left( {\delta _ - ^2} \right) = \frac{1}{{2\left( {\delta _ - ^2 + {\varepsilon _i}} \right)}}\cdot\left[ {{\varepsilon _i} \pm \sqrt {{\varepsilon _i}\left( {{\varepsilon _i} - 1} \right) - \delta _ - ^2} } \right].
\end{array}
\end{equation}
Where
\begin{equation} \label{eq:33}
\begin{aligned}
{x_{\left( j \right)}} = \frac{{{E_{\left( j \right)}}}}{{{\omega _i}}},\quad j = a+,{\rm{ }}b- \\
{\varepsilon _i} = \frac{{{\omega _i}}}{{{\omega _{thr}}}},\quad {\omega _{thr}} = \frac{{{m^2}}}{{\omega {{\sin }^2}\left( {{{{\theta _i}} \mathord{\left/
 {\vphantom {{{\theta _i}} 2}} \right.
 \kern-\nulldelimiterspace} 2}} \right)}}
\end{aligned}
\end{equation}
Note that the energies of electron (channel a) and positron (channel b) might be obtained from the energy conservation law ${x_{\left( a \right) - }}\left( {\delta _ + ^2} \right) \approx 1 - {x_{\left( a \right) + }}\left( {\delta _ + ^2} \right)$, ${x_{\left( b \right) + }}\left( {\delta _ - ^2} \right) \approx 1 - {x_{\left( b \right) - }}\left( {\delta _ - ^2} \right)$.

From the relations (31) and (32) follows that the minimum energy of the initial gamma-quantum at resonance will be at ${\varepsilon _{i\min }} = 1$ and $\delta _ \pm ^2 = 0$, i.e. ${\omega _{thr}}$ is the threshold energy. The threshold energy is determined by the electron's rest energy, the frequency of the electromagnetic wave and the angle between the momenta of the initial photon and the electromagnetic wave. For the frequencies from the optical range the threshold energy is of the order of magnitude ${\omega _{thr}} \sim {10^5} \div {10^6}\;{\rm{MeV}}$, and for x-ray laser  ${\omega _{thr}} \sim {10^2} \div {10^3}\;{\rm{MeV}}$.

Thus, the resonant energies of a positron and an electron are determined by two parameters: parameter ${\varepsilon _i}$ is the energy of the initial gamma-quantum in units of the threshold energy, and also ultrarelativistic parameters $\delta _ \pm ^2$ (30) determining possible radiation angles of a positron and an electron. It is important to emphasize that the resonant energies of the positron and the electron for channel a) are determined only by the radiation angle of the positron $\left( {\delta _ + ^2} \right)$ and for channel b) only by electron radiation angle $\left( {\delta _ - ^2} \right)$. Moreover, the resonant energy of the electron-positron pair can take two different values for each angle. Possible values of angles of the pair substantially depend on the value of the parameter ${\varepsilon _i}$ and are enclosed in the interval
\begin{equation} \label{eq:34}
0 \le \delta _ \pm ^2 \le {\varepsilon _i}\left( {{\varepsilon _i} - 1} \right),\quad {\varepsilon _i} \ge 1
\end{equation}

This shows that if the energy of the initial gamma-quantum is equal to the threshold energy $\left( {{\varepsilon _i} = 1} \right)$ then the electron-positron pair propagates exactly along the momentum of the initial photon $\left( {\delta _ \pm ^2 = 0} \right)$ and the possible two values of the energy of the positron (electron) differ as much as possible from each other. With the increase of the radiation angles of the electron-positron pair due to the relation (34) the difference in the two possible energies of the positron (electron) decreases. With the maximum possible radiation angle $\delta _{ \pm \max }^2 = {\varepsilon _i}\left( {{\varepsilon _i} - 1} \right)$ the energy of the positron (electron) takes a single value (see (31), (32), also Fig.3.1 and Fig.3.2). Note that in the papers [12-14, 18, 19, 27] resonances were studied only for channel a), when the positron propagates along the initial photon momentum $\left( {\delta _ + ^2 = 0} \right)$, and the electron is scattered on the nucleus at large angles. Here we study the resonances for channels a) and b) in the case of high energies of the initial photon (${\varepsilon _i} \ge 1\;\left( {{\omega _i} \ge {\omega _{thr}}} \right)$), when the ultrarelativistic electron-positron pair propagates in a narrow cone along the momentum of the initial photon. The interference of the channels under the resonant conditions will occur when the energies of the positron (electron) for channels a) and b) are equal (${E_{ + \left( a \right)}} = {E_{ + \left( b \right)}}$) (see Fig. 3.1 and Fig. 3.2).  If we exclude these cases, the interference of channels a) and b) in the conditions of resonance will not take place. This case will be considered in this paper, when 
\begin{equation} \label{eq:35}
{E_{ + \left( a \right)}} \ne {E_{ + \left( b \right)}}\quad \left( {{E_{ - \left( a \right)}} \ne {E_{ - \left( b \right)}}} \right)
\end{equation}
\begin{figure}[h!]
\begin{minipage}[h!]{1.0\linewidth}
     \includegraphics[width=10cm]{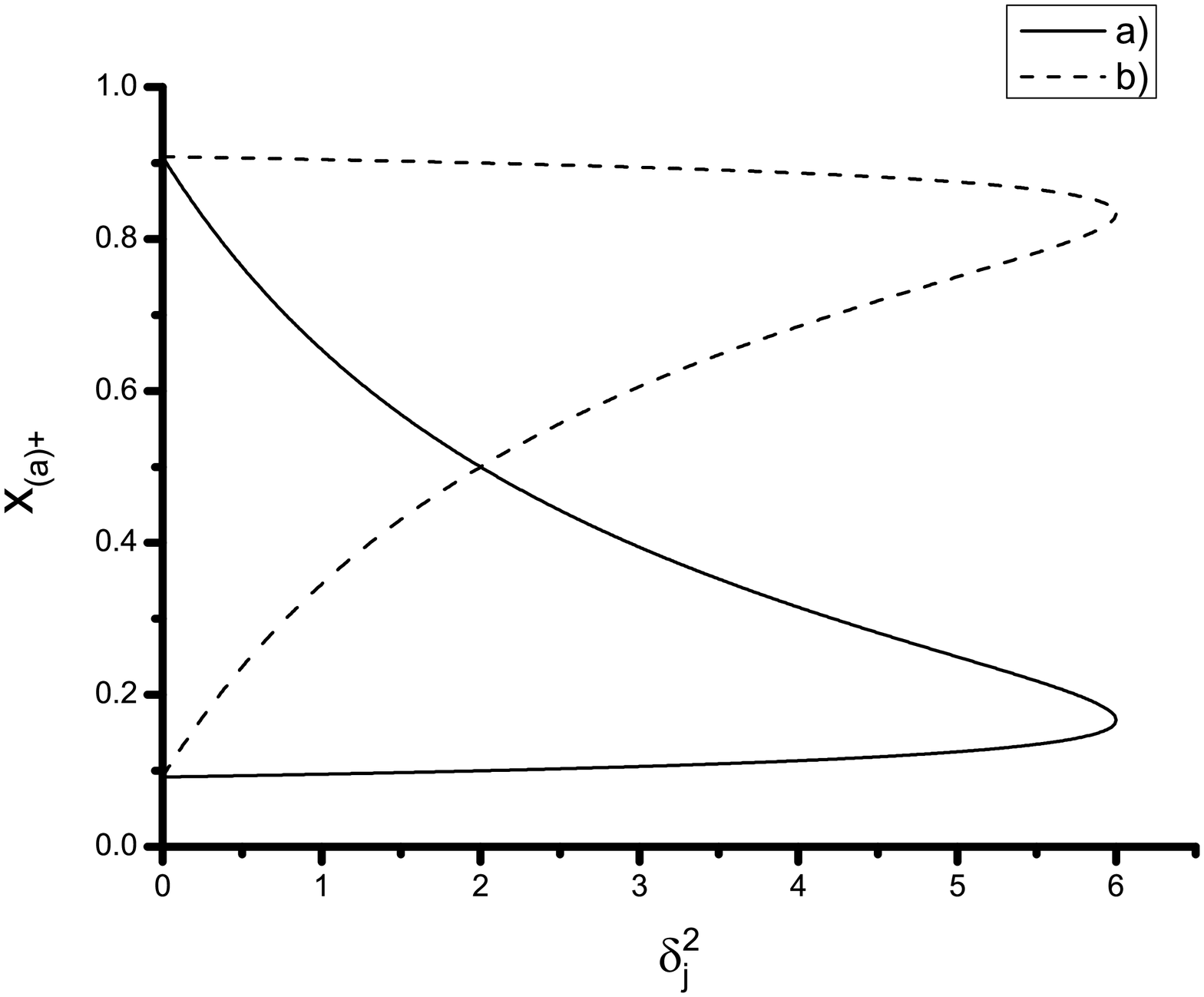}
\caption*{FIG. 3.1: Dependence of positron resonant energy on ultrarelativistic parameter (in units of energy of initial photon):$\delta _j^2 = \delta _ + ^2$  - for channel a) (solid line), $\delta _j^2 = \delta _ - ^2$ - for channel b) (dashed line). Parameter ${\varepsilon _i} = 3$.}
\end{minipage}
\begin{minipage}{1.0\linewidth}
     \includegraphics[width=10cm]{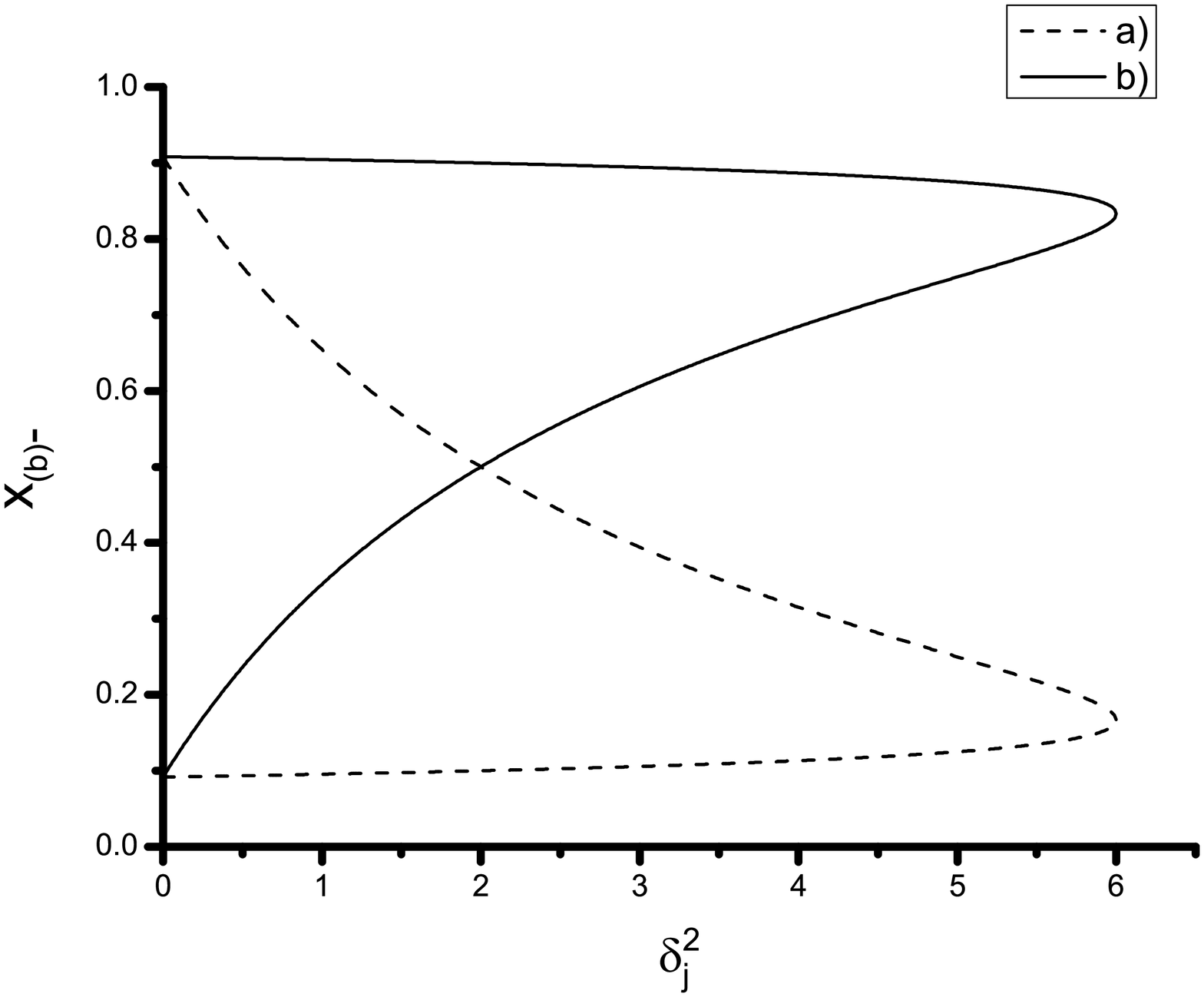}
     \caption* {FIG. 3.2: Dependence of electron resonant energy on ultrarelativistic parameter (in units of energy of initial photon): $\delta _j^2 = \delta _ - ^2$ - for channel b) (solid line), $\delta _j^2 = \delta _ + ^2$ - for channel a) (dashed line). Parameter ${\varepsilon _i} = 3$.}
\end{minipage}
  \end{figure}
\section{\label{sec:level1}The resonant differential cross section}
Simplify the amplitude of the process of resonant PPP on the nucleus in the field of a plane electromagnetic wave. For partial amplitude (6),(7),(12),(13) in a weak field (3) with ${l_1} = 1$ obtain:
\begin{equation} \label{eq:36}
{S_l} = i \cdot \frac{{8Z{e^3}{\pi ^{5/2}}}}{{\sqrt {2{E_ - }{E_ + }{\omega _i}} }} \cdot \exp (i\tilde \phi ) \cdot \left[ {\bar u{M_l}v} \right] \cdot \frac{{\delta ({q_0})}}{{{{\bf{q}}^2}}},
\end{equation}

Expression for the differential cross section was obtained by a standard procedure (see [48]). After simple calculations we have:
\begin{equation} \label{eq:37}
d\sigma  = \sum\limits_{l =  - \infty }^\infty  {d{\sigma _l}} ,
\end{equation}
Where the partial differential cross section is:
\begin{equation} \label{eq:38}
d{\sigma _l} = \frac{{{{\left( {Z{e^3}} \right)}^2}}}{{2\pi {\omega _i}{E_ + }{E_ - }}}\frac{{{{\left| {\bar u{M_l}v} \right|}^2}}}{{{{\bf{q}}^4}}}\delta \left( {{q_0}} \right){d^3}{p_ + }{d^3}{p_ - },
\end{equation}
\begin{equation} \label{eq:39}
{M_l} = \frac{{{L_{l + 1}}({p_ - },{q_ - })}}{{q_ - ^2 - {m^2}}}\left( {{\varepsilon _\mu }G_ - ^\mu } \right) + \frac{{{L_{l + 1}}({p_ + },{q_ + })}}{{q_ + ^2 - {m^2}}}\left( {{\varepsilon _\mu }G_ + ^\mu } \right),
\end{equation}
\begin{equation} \label{eq:40}
G_ - ^\mu  = {\tilde \gamma ^0}({\hat q_ - } + m)Y_ - ^\mu ,{\rm{  }}G_ + ^\mu  = Y_ + ^\mu ({\hat q_ + } + m){\tilde \gamma ^0}
\end{equation}
Here matrices $Y_ - ^\mu \;{\rm{ and }}\;Y_ + ^\mu $ are equal to:
\begin{equation} \label{eq:41}
\begin{aligned}
Y_ - ^\mu  = {{\tilde \gamma }^\mu } \cdot {a_0}\left( {{q_ - },{p_ + }} \right) + {a_1}\left( {{q_ - },{p_ + }} \right) \cdot \left( {{k^\mu }{{\hat \varepsilon }_ - } - \varepsilon _ - ^\mu \hat k} \right) + \\
 + {a_2}\left( {{q_ - },{p_ + }} \right) \cdot {{\hat \varepsilon }_ - }\hat k{{\tilde \gamma }^\mu },
\end{aligned}
\end{equation}
\begin{equation} \label{eq:42}
\begin{aligned}
\;{a_0} = {L_1}\left( {{q_ - },{p_ + }} \right),\quad {a_1} = \eta \frac{m}{{2\left( {k{p_ + }} \right)}}\cdot{L_0}\left( {{q_ - },{p_ + }} \right),\\
{a_2} =  - \frac{1}{4}\eta m \cdot \left[ {\frac{1}{{\left( {k{p_ + }} \right)}} + \frac{1}{{\left( {k{q_ - }} \right)}}} \right] \cdot {L_0}\left( {{q_ - },{p_ + }} \right)
\end{aligned}
\end{equation}
and
\begin{equation} \label{eq:43}
\begin{aligned}
Y_ + ^\mu  = {{\tilde \gamma }^\mu } \cdot {a_0}\left( {{p_ - },{q_ + }} \right) + {a_1}\left( {{p_ - },{q_ + }} \right) \cdot \left( {{k^\mu }{{\hat \varepsilon }_ - } - \varepsilon _ - ^\mu \hat k} \right) + \\
 + {a_2}\left( {{p_ - },{q_ + }} \right) \cdot {{\hat \varepsilon }_ - }\hat k{{\tilde \gamma }^\mu },
\end{aligned}
\end{equation}
\begin{equation} \label{eq:44}
\begin{aligned}
{a_0} = {L_1}\left( {{p_ - },{q_ + }} \right),\quad {a_1} = \eta \frac{m}{{2\left( {k{p_ - }} \right)}}\cdot{L_0}\left( {{p_ - },{q_ + }} \right), \\
{a_2} =  - \frac{1}{4}\eta m \cdot \left[ {\frac{1}{{\left( {k{p_ - }} \right)}} + \frac{1}{{\left( {k{q_ + }} \right)}}} \right] \cdot {L_0}\left( {{p_ - },{q_ + }} \right).
\end{aligned}
\end{equation}

Obtain the cross section of the resonant process in the absence of the interference of amplitude (first and second terms in (39)). We first calculate the resonant cross section for channel a)
\begin{equation} \label{eq:45}
d{\sigma _l} = \frac{{{{\left( {Z{e^3}} \right)}^2}}}{{2\pi {\omega _i}{E_ + }{E_ - }}}\frac{{{{\left| {{L_{l + 1}}({p_ - },{q_ - })} \right|}^2}}}{{{{\left| {q_ - ^2 - {m^2}} \right|}^2}{{\bf{q}}^4}}}{\left| {\bar u{\varepsilon _\mu }G_ - ^\mu v} \right|^2}\delta \left( {{q_0}} \right){d^3}{p_ + }{d^3}{p_ - },
\end{equation}
We consider non-polarized particles. Then the averaging over the polarizations of the initial photons and the summation over the polarizations of the electrons and positrons in the final state are reduced to the replacement:
\begin{equation} \label{eq:46}
\!\:{\left| {\bar u{\varepsilon _\mu }G_ - ^\mu v} \right|^2} \to  - \frac{1}{2}Tr\left[ {{G_{ - \mu }}\left( {{{\hat p}_ + } - m} \right)\bar G_ - ^\mu \left( {{{\hat p}_ - } + m} \right)} \right]
\end{equation}
After calculations of the trace (46) let us integrate the resulting expression over the energy of the final electron using the Dirac delta function $\delta \left( {{q_0}} \right)$. As a result, the resonant differential cross section of the PPP at the nucleus in the field of the plane electromagnetic wave for channels a)  $(d{\sigma _{\left( a \right)res}})$ and b) $(d{\sigma _{\left( b \right)res}})$ takes the form:
\begin{equation} \label{eq:47}
d{\sigma _{res}} = d{\sigma _{\left( j \right)res}},\quad j = a,b
\end{equation}
where
\begin{equation} \label{eq:48}
d{\sigma _{\left( a \right)res}} = d{\sigma _{l + 1}}\left( {{p_ - },{q_ - }} \right) \cdot \frac{{{m^2}\left| {{{\bf{q}}_ - }} \right|}}{{2\pi {{\left| {q_ - ^2 - {m^2}} \right|}^2}}} \cdot d{W_1}\left( {{q_ - },{p_ + }} \right),
\end{equation}
\begin{equation} \label{eq:49}
\begin{aligned}
d{W_1}\left( {{q_ - },{p_ + }} \right) = \frac{\alpha }{{{\omega _i}{E_ + }}}{\eta ^2}\left[ {\frac{{\gamma _{{p_ + }{q_ - }}^2}}{{{\eta ^2}}} + } \right.\\
 + \left( {1 + {\delta ^2}} \right)\left( {2{u_{(a)}} - 1} \right)\Bigg]{d^3}{p_ + },
\end{aligned}
\end{equation}
\begin{equation} \label{eq:50}
\begin{aligned}
d{\sigma _{l + 1}}\left( {{p_ - },{q_ - }} \right) = 2{Z^2}r_e^2\frac{{\left| {{{\bf{p}}_ - }} \right|}}{{\left| {{{\bf{q}}_ - }} \right|}}\frac{{{m^2}\left( {{m^2} + p_ - ^0q_ - ^0 + {{\bf{p}}_ - }{{\bf{q}}_ - }} \right)}}{{{{\left( {{{\bf{p}}_ - } + {{\bf{p}}_ + } - {{\bf{k}}_i} - \left( {l + 1} \right){\bf{k}}} \right)}^4}}} \cdot \\
 \cdot {\left| {{L_{l + 1}}\left( {{p_ - },{q_ - }} \right)} \right|^2}d{\Omega _ -},
\end{aligned}
\end{equation}
\begin{equation} \label{eq:51}
d{\sigma _{\left( b \right)res}} = d{W_1}\left( {{p_ - },{q_ + }} \right) \cdot \frac{{{m^2}\left| {{{\bf{q}}_ + }} \right|}}{{2\pi {{\left| {q_ + ^2 - {m^2}} \right|}^2}}} \cdot d{\sigma _{l + 1}}\left( {{q_ + },{p_ + }} \right),
\end{equation}
\begin{equation} \label{eq:52}
\begin{aligned}
d{W_1}\left( {{p_ - },{q_ + }} \right) = \frac{\alpha }{{{\omega _i}{E_ - }}}{\eta ^2}\left[ {\frac{{\gamma _{{p_ - }{q_ + }}^2}}{{{\eta ^2}}} + } \right.\\
 + \left( {1 + {\delta ^2}} \right)\left( {2{u_{(b)}} - 1} \right)\Bigg]{d^3}{p_ - },
\end{aligned}
\end{equation}
\begin{equation} \label{eq:53}
\begin{aligned}
d{\sigma _{l + 1}}\left( {{q_ + },{p_ + }} \right) = 2{Z^2}r_e^2\frac{{\left| {{{\bf{p}}_ + }} \right|}}{{\left| {{{\bf{q}}_ + }} \right|}}\frac{{{m^2}\left( {{m^2} + p_ + ^0q_ + ^0 + {{\bf{p}}_ + }{{\bf{q}}_ + }} \right)}}{{{{\left( {{{\bf{p}}_ - } + {{\bf{p}}_ + } - {{\bf{k}}_i} - \left( {l + 1} \right){\bf{k}}} \right)}^4}}} \cdot \\
 \cdot {\left| {{L_{l + 1}}\left( {{q_ + },{p_ + }} \right)} \right|^2}d{\Omega _ +},
\end{aligned}
\end{equation}

Here in relations (50) and (53 ) solid angles of positron and electron are denoted by $d{\Omega _ \pm }$. In expressions (49) and (52) parameters ${\gamma _{{p_ + }{q_ - }}}$ and ${\gamma _{{p_ - }{q_ + }}}$ are determined by relations (18), (19), where it is necessary to replace: ${p_1} \to {p_ + },{\rm{  }}{p_2} \to {q_ - }$ for ${\gamma _{{p_ + }{q_ - }}}$ and ${p_1} \to {q_ + },{\rm{  }}{p_2} \to {p_ - }$ for ${\gamma _{{p_ - }{q_ + }}}$. Relativistic-invariant parameters ${u_{(a)}}\;{\rm{ and }}\;{u_{(b)}}$  equals to:
\begin{equation} \label{eq:54}
{u_{(a)}} = \frac{{{{(k{k_i})}^2}}}{{4(k{q_ - })(k{p_ + })}};\quad{u_{(b)}} = \frac{{{{(k{k_i})}^2}}}{{4(k{q_ + })(k{p_ - })}}
\end{equation}

From the relations (48)-(53) we can see that for channels a) and b), the resonant differential cross section of the PPP at the nucleus in the field of the plane electromagnetic wave effectively reduce into two first-order processes in the fine structure constant. For the channel a) first takes a place laser-stimulated Breit-Wheeler process ($d{W_1}\left( {{q_ - },{p_ + }} \right)$ - is the probability of this process per unit of time [15, 48]), and then laser-assisted process of scattering of the intermediate electron on the nucleus ($d\sigma \left( {{p_ - },{q_ - }} \right)$ - is the differential cross section of this process [8, 11]). A similar situation for channel b). However here we have laser-assisted process of scattering of the intermediate positron on the nucleus ($d\sigma \left( {{q_ + },{p_ + }} \right)$ - is the differential cross section of this process).

Transform relativistic resonant cross sections (48) and (51) into resonant kinematics (27)-(30). In this case, we consider that for scattering of an electron or a positron on the nucleus at small angles, the following relations take a place [44]
\begin{equation} \label{eq:55}
\begin{aligned}
{\left| {{L_{l + 1}}\left( {{p_ - },{q_ - }} \right)} \right|^2} = J_{l + 1}^2\left( {{\gamma _{{p_ - }{q_ - }}}} \right) \approx 1\\
{\left| {{L_{l + 1}}\left( {{q_ + },{p_ + }} \right)} \right|^2} = J_{l + 1}^2\left( {{\gamma _{{q_ + }{p_ + }}}} \right) \approx 1
\end{aligned}
\end{equation}
Since $\left( {{\gamma _{{p_ - }{q_ - }}}\sim\eta \ll1} \right)$  and $\left( {{\gamma _{{q_ + }{p_ + }}}\sim\eta \ll 1} \right)$ in the process of scattering of an electron or positron on the nucleus, the process with absorption of a one photon from electromagnetic wave is more probable, i.e. $l =  - 1$ .

Further we consider circular polarization of the electromagnetic wave $\left( {{\delta ^2} = 1} \right)$. Elimination of the resonant infinity in the channels a) and b) can be accomplished by adding of imaginary term to the mass of the intermediate electron and positron. So, for channel a) we have:
\begin{equation} \label{eq:56}
m \to \mu  = m + i{\Gamma _{\left( a \right)}},\quad {\Gamma _{\left( a \right)}} = \frac{{{q^0_{-}}}}{{2m}}{W_{\left( a \right)}}.
\end{equation}
Here ${W_{\left( a \right)}}$ is the total probability (per unit time) of laser-stimulated Breit-Wheeler process [15].
\begin{equation} \label{eq:57}
{W_{\left( a \right)}} = \frac{{\alpha {m^2}}}{{8\pi {\omega _i}}}{\eta ^2}{K_i}
\end{equation}
\begin{equation} \label{eq:58}
\begin{aligned}
{K_i} = \left( {2 + \frac{2}{{{\varepsilon _i}}} - \frac{1}{{\varepsilon _i^2}}} \right){\rm{Artanh}}\left( {\sqrt {\frac{{{\varepsilon _i} - 1}}{{{\varepsilon _i}}}} } \right) - \\
 - \left( {\frac{{{\varepsilon _i} + 1}}{{{\varepsilon _i}}}} \right)\sqrt {\frac{{{\varepsilon _i} - 1}}{{{\varepsilon _i}}}} 
\end{aligned}
\end{equation}
Taking into account the relations (56), the resonant denominator can be represented as:
\begin{equation} \label{eq:59}
{\left| {q_ - ^2 - {\mu ^2}} \right|^2} = 16{m^4}\left[ {x_{(a) + }^2\left( {\delta _ + ^2 - \delta _{(a) + }^2} \right) + \frac{{4{\Gamma^2 _{(a)}}}}{{{m^2}}}} \right]
\end{equation}
the parameter $\delta _{(a)+}^2$ is related to the resonant frequency of the incident gamma-quantum for channel a) by the expression:
\begin{equation} \label{eq:60}
\delta _{(a) + }^2 = \frac{{4{\varepsilon _i}{x_{(a) + }}\left( {1 - {x_{(a) + }}} \right) - 1}}{{4x_{(a) + }^2}}
\end{equation}
Similarly, we can define the expression for the resonant denominator of the channel b):
\begin{equation} \label{eq:61}
{\left| {q_ + ^2 - {\mu ^2}} \right|^2} = 16{m^4}\left[ {x_{(b) - }^2\left( {\delta _ - ^2 - \delta _{(b) - }^2} \right) + \frac{{4{\Gamma^2 _{(b)}}}}{{{m^2}}}} \right]
\end{equation}
the parameter $\delta _{(b)-}^2$ is related to the resonant frequency of the incident gamma-quantum for channel a) by the expression:
\begin{equation} \label{eq:62}
\delta _{(b) - }^2 = \frac{{4{\varepsilon _i}{x_{(b) - }}\left( {1 - {x_{(b) - }}} \right) - 1}}{{4x_{(b) - }^2}}
\end{equation}

After the simple transformations of (48) and (51) obtain following expressions for the resonant differential cross sections of PPP in the case of an ultrarelativistic pair:
\begin{equation} \label{eq:63}
\begin{aligned}
d{\sigma _{\left( a \right)res}} = \frac{{{Z^2}{\eta ^2}\alpha r_e^2}}{{{{\left[ {d\left( {{x_{\left( a \right) + }}} \right)} \right]}^2}}}\frac{{{{\left( {1 - {x_{\left( a \right) + }}} \right)}^3} \cdot G\left( {{x_{\left( a \right) + }}} \right)}}{{\left[ {{{\left( {\delta _ + ^2 - \delta _{\left( a \right) + }^2} \right)}^2} + \Gamma _{{\delta _ + }}^2} \right]}} \cdot \\
 \cdot \frac{{d{x_{\left( a \right) + }}}}{{{x_{\left( a \right) + }}}}d\delta _ + ^2d\delta _ - ^2d\varphi ,
\end{aligned}
\end{equation}
\begin{equation} \label{eq:64}
\begin{aligned}
d{\sigma _{\left( b \right)res}} = \frac{{{Z^2}{\eta ^2}\alpha r_e^2}}{{{{\left[ {d\left( {{x_{\left( b \right) - }}} \right)} \right]}^2}}}\frac{{{{\left( {1 - {x_{\left( b \right) - }}} \right)}^3} \cdot G\left( {{x_{\left( b \right) - }}} \right)}}{{\left[ {{{\left( {\delta _ - ^2 - \delta _{\left( b \right) - }^2} \right)}^2} + \Gamma _{{\delta _ - }}^2} \right]}} \cdot \\
 \cdot \frac{{d{x_{\left( b \right) - }}}}{{{x_{\left( b \right) - }}}}d\delta _ - ^2d\delta _ + ^2d\varphi ,
\end{aligned}
\end{equation}
Here $\varphi$ - is an angle between planes $\left( {{{\bf{k}}_i},{{\bf{p}}_ + }} \right)$ and $\left( {{{\bf{k}}_i},{{\bf{p}}_ - }} \right)$. ${\Gamma _{{\delta _ + }}}$ and ${\Gamma _{{\delta _ - }}}$ - are the angular radiative resonance width for channels a) and b)
\begin{equation} \label{eq:65}
\Gamma _{{\delta _ + }} = \frac{{\alpha {\eta ^2}}}{{32\pi }}\frac{{\left( {1 - {x_{(a) + }}} \right)}}{{{x_{(a) + }}}}{K_i};\;\Gamma _{{\delta _ - }}= \frac{{\alpha {\eta ^2}}}{{32\pi }}\frac{{\left( {1 - {x_{(b) - }}} \right)}}{{{x_{(b) - }}}}{K_i}
\end{equation}
\begin{equation} \label{eq:66}
G\left( {{x_ \pm }} \right) = \frac{{4\tilde \delta _ \pm ^2}}{{{{\left( {1 + \tilde \delta _ \pm ^2} \right)}^2}}} + \left( {\frac{{{x_ \pm }}}{{1 - {x_ \pm }}} + \frac{{1 - {x_ \pm }}}{{{x_ \pm }}}} \right),
\end{equation}
\begin{equation} \label{eq:67}
\begin{aligned}
d\left( {{x_ \pm }} \right) = {d_0}\left( {{x_ \pm }} \right) + {\left( {\frac{m}{{2{\omega _i}}}} \right)^2}\cdot\biggl[d_1^2\left( {{x_ \pm }} \right) + \\
 + \left. {\frac{{4{\varepsilon _i}}}{{\sin \left( {{{{\theta _i}} \mathord{\left/
 {\vphantom {{{\theta _i}} 2}} \right.
 \kern-\nulldelimiterspace} 2}} \right)}}\left\{ {4{\varepsilon _i} - {d_1}\left( {{x_ \pm }} \right)} \right\}} \right]
\end{aligned}
\end{equation}
\begin{equation} \label{eq:68}
{d_0}\left( {{x_ \pm }} \right) = \tilde \delta _ + ^2 + \tilde \delta _ - ^2 + 2\tilde \delta _ + ^{}\tilde \delta _ - ^{}\cos \varphi 
\end{equation}
\begin{equation} \label{eq:69}
{d_1}\left( {{x_ + }} \right) = \frac{{1 + \tilde \delta _ + ^2}}{{{x_ + }}} + \frac{{1 + \tilde \delta _ - ^2}}{{1 - {x_ + }}};\;{d_1}\left( {{x_ - }} \right) = \frac{{1 + \tilde \delta _ - ^2}}{{{x_ - }}} + \frac{{1 + \tilde \delta _ + ^2}}{{1 - {x_ - }}}
\end{equation}

For the same energies of the positron and the electron for the resonant process in a wave field, the differential cross section of the PPP without an external field $d{\sigma _0}$ has form [48]:
\begin{equation} \label{eq:70}
\begin{aligned}
d{\sigma _0} = \frac{{128}}{\pi }{Z^2}r_e^2\alpha x_ \pm ^3{\left( {1 - {x_ \pm }} \right)^3}\frac{{{M_0} + {{\left( {m/{\omega _i}} \right)}^2}M}}{{{{\left[ {{d_0} + {{\left( {m/2{\omega _i}} \right)}^2}d_1^2} \right]}^2}}} \cdot \\
 \cdot d\delta _ + ^2d\delta _ - ^2d{x_ \pm }d\varphi 
\end{aligned}
\end{equation}
\begin{equation} \label{eq:71}
\begin{aligned}
{M_0}\left( {{x_ \pm }} \right) =  - \frac{{\tilde \delta _ + ^2}}{{{{\left( {1 + \tilde \delta _ + ^2} \right)}^2}}} - \frac{{\tilde \delta _ - ^2}}{{{{\left( {1 + \tilde \delta _ - ^2} \right)}^2}}} + \\
 + \frac{1}{{2{x_ \pm }(1 - {x_ \pm })}}\frac{{\tilde \delta _ + ^2 + \tilde \delta _ - ^2}}{{\left( {1 + \tilde \delta _ + ^2} \right)\left( {1 + \tilde \delta _ - ^2} \right)}} + \\
 + \left( {\frac{{{x_ \pm }}}{{1 - {x_ \pm }}} + \frac{{1 - {x_ \pm }}}{{{x_ \pm }}}} \right)\frac{{{{\tilde \delta }_ + }{{\tilde \delta }_ - }\cos \varphi }}{{\left( {1 + \tilde \delta _ + ^2} \right)\left( {1 + \tilde \delta _ - ^2} \right)}},
\end{aligned}
\end{equation}
\begin{equation} \label{eq:72}
M\left( {{x_ \pm }} \right) = \left( {\frac{1}{{x_ \pm ^2}} + \frac{1}{{{{\left( {1 - {x_ \pm }} \right)}^2}}}} \right){b_ \pm },
\end{equation}
\begin{equation} \label{eq:73}
\begin{aligned}
{b_ \pm }\left( {{x_ \pm }} \right) = \frac{{\tilde \delta _ \pm ^2}}{{12{{\left( {1 + \tilde \delta _ \pm ^2} \right)}^3}}}\biggl[2\left( {1 - \tilde \delta _ \pm ^2} \right)\left( {3 - \tilde \delta _ \pm ^2} \right)-\\
 - \frac{1}{{{x_ \pm }(1 - {x_ \pm })}}\left( {9 + 2\tilde \delta _ \pm ^2 + \tilde \delta _ \pm ^4} \right)\\
\left. { + \left[ {\frac{{{x_ \pm }}}{{1 - {x_ \pm }}} + \frac{{1 - {x_ \pm }}}{{{x_ \pm }}}} \right]\left( {9 + 4\tilde \delta _ \pm ^2 + 3\tilde \delta _ \pm ^4} \right)} \right\}.
\end{aligned}
\end{equation}

Note that in the formulas for the differential cross section of the PPP without a laser field (70)-(73)   for channel a) must select the plus sign $\left( {{x_ + }} \right)$ and for channel b) the minus sign $\left( {{x_ - }} \right)$. It is important to emphasize that in differential cross sections (63),(64) and (70) introduced small corrections proportional to the value $ \sim {\left( {m/{\omega _i}} \right)^2} \ll 1$ . We note that these corrections make the dominant contribution to the corresponding differential cross sections under conditions:
\begin{equation} \label{eq:74}
\left| {\varphi  - \pi } \right| \mathbin{\lower.3ex\hbox{$\buildrel<\over
{\smash{\scriptstyle\sim}\vphantom{_x}}$}} \frac{m}{{{\omega _i}}},{\rm{  }}\left| {{{\tilde \delta }_ + } - {{\tilde \delta }_ - }} \right| \mathbin{\lower.3ex\hbox{$\buildrel<\over
{\smash{\scriptstyle\sim}\vphantom{_x}}$}} \frac{m}{{{\omega _i}}}
\end{equation}
In this case, the values ${M_0} \to 0\;\rm{and}\;{d_0} \to 0$ and corresponding differential cross sections has sharp maximum. So, the differential cross sections without field (70) and in the field of wave (63),(64) in the kinematic range (74) will have the following order of magnitude:
\begin{equation} \label{eq:75}
d{\sigma _0} \sim {Z^2}\alpha r_e^2{\left( {\frac{{{\omega _i}}}{m}} \right)^2},\;d{\sigma _{\left( a \right)res}} \mathbin{\lower.3ex\hbox{$\buildrel<\over
{\smash{\scriptstyle\sim}\vphantom{_x}}$}} \frac{{{Z^2}\alpha r_e^2}}{{\left(\alpha \eta\right)^2}}{\left( {\frac{{{\omega _i}}}{m}} \right)^4}
\end{equation}
We integrate the resonant differential cross sections, as well as cross section in the absence of the field with the respect to the azimuth angle $\varphi$. After simple calculations, we obtain:
\begin{equation} \label{eq:76}
\begin{aligned}
d\sigma _{(a)res}^{} = 2\pi \left( {{Z^2}\alpha r_e^2} \right){\eta ^2}\frac{{{{\left( {1 - {x_{(a) + }}} \right)}^3}}}{{{x_{(a) + }}}} \cdot \\
 \cdot \frac{{{H_{1( + )}}G({x_{(a) + }})}}{{\left[ {{{\left( {\delta _ + ^2 - \delta _{\left( a \right) + }^2} \right)}^2} + \Gamma _{{\delta _ + }}^2} \right]}} d{x_{(a) + }}d\delta _ + ^2d\delta _ - ^2,
\end{aligned}
\end{equation}
\begin{equation} \label{eq:77}
\begin{aligned}
d\sigma _{(b)res}^{} = 2\pi \left( {{Z^2}\alpha r_e^2} \right){\eta ^2}\frac{{{{\left( {1 - {x_{(b) - }}} \right)}^3}}}{{{x_{(b) - }}}} \cdot \\
 \cdot \frac{{{H_{1( - )}}G({x_{(b) - }})}}{{\left[ {{{\left( {\delta _ - ^2 - \delta _{\left( b \right) - }^2} \right)}^2} + \Gamma _{{\delta _ - }}^2} \right]}}d{x_{(b) - }}d\delta _ + ^2d\delta _ - ^2,
\end{aligned}
\end{equation}
\begin{equation} \label{eq:78}
d{\sigma _0} = 64\left( {{Z^2}\alpha r_e^2} \right)x_ \pm ^3{(1 - {x_ \pm })^3}{H_0} \cdot d{x_ \pm }d\delta _ + ^2d\delta _ - ^2,
\end{equation}
\begin{equation} \label{eq:79}
{H_0} = \frac{{\left( {\tilde \delta _ + ^2 + \tilde \delta _ - ^2} \right)\left[ {{D_0} + {{\left( {m/{\omega _i}} \right)}^2}D} \right]}}{{{{\left[ {{{\left( {\tilde \delta _ + ^2 - \tilde \delta _ - ^2} \right)}^2} + \frac{1}{2}{{\left( {\frac{m}{{{\omega _i}}}} \right)}^2}\left( {\tilde \delta _ + ^2 + \tilde \delta _ - ^2} \right)d_1^2} \right]}^{3/2}}}},
\end{equation}
\begin{equation} \label{eq:80}
\begin{aligned}
D_0 =  - \frac{{\tilde \delta _ + ^2}}{{{{\left( {1 + \tilde \delta _ + ^2} \right)}^2}}} - \frac{{\tilde \delta _ - ^2}}{{{{\left( {1 + \tilde \delta _ - ^2} \right)}^2}}} + \\
 + \frac{1}{{2{x_ \pm }(1 - {x_ \pm })}}\frac{{\tilde \delta _ + ^2 + \tilde \delta _ - ^2}}{{\left( {1 + \tilde \delta _ + ^2} \right)\left( {1 + \tilde \delta _ - ^2} \right)}} + \\
 + \left( {\frac{{{x_ \pm }}}{{1 - {x_ \pm }}} + \frac{{1 - {x_ \pm }}}{{{x_ \pm }}}} \right)\frac{{2\tilde \delta _ + ^2\tilde \delta _ - ^2}}{{\left( {\tilde \delta _ + ^2 + \tilde \delta _ - ^2} \right)\left( {1 + \tilde \delta _ + ^2} \right)\left( {1 + \tilde \delta _ - ^2} \right)}},
\end{aligned}
\end{equation}
\begin{equation} \label{eq:81}
\begin{aligned}
D = M\left( {{x_ \pm }} \right) + d_1^2\left( {{x_ \pm }} \right) \cdot \left( {\frac{{{x_ \pm }}}{{1 - {x_ \pm }}} + \frac{{1 - {x_ \pm }}}{{{x_ \pm }}}} \right) \cdot \\
 \cdot \frac{{\tilde \delta _ + ^2\tilde \delta _ - ^2}}{{2\left( {\tilde \delta _ + ^2 + \tilde \delta _ - ^2} \right)\left( {1 + \tilde \delta _ + ^2} \right)\left( {1 + \tilde \delta _ - ^2} \right)}},
\end{aligned}
\end{equation}
\begin{equation} \label{eq:82}
\begin{aligned}
{H_{1( \pm )}} = \left( {\tilde \delta _ + ^2 + \tilde \delta _ - ^2} \right)\left\{ {{{\left( {\tilde \delta _ + ^2 - \tilde \delta _ - ^2} \right)}^2} + \frac{1}{2}{{\left( {\frac{m}{{{\omega _i}}}} \right)}^2}\left( {\tilde \delta _ + ^2 + \tilde \delta _ - ^2} \right)} \right. \cdot \\
{\left. { \cdot \left[ {d_1^2\left( {{x_ \pm }} \right) + \frac{{4{\varepsilon _i}}}{{\sin \left( {{{{\theta _i}} \mathord{\left/
 {\vphantom {{{\theta _i}} 2}} \right.
 \kern-\nulldelimiterspace} 2}} \right)}}\left\{ {4{\varepsilon _i} - {d_1}\left( {{x_ \pm }} \right)} \right\}} \right]} \right\}^{ - 3/2}}.
\end{aligned}
\end{equation}
It is important to emphasize that under the condition
\begin{equation} \label{eq:83}
\left| {\tilde \delta _ + ^2 - \tilde \delta _ - ^2} \right| \mathbin{\lower.3ex\hbox{$\buildrel<\over
{\smash{\scriptstyle\sim}\vphantom{_x}}$}} \frac{m}{{{\omega _i}}}
\end{equation}
the resonant cross sections (76), (77), and also differential cross section of PPP in the absence of the laser field (78) have sharp maximum, which is associated with small angels of scattering. So we need to take into account small relativistic corrections $ \sim {\left( {{m \mathord{\left/
 {\vphantom {m {{\omega _i}}}} \right.
 \kern-\nulldelimiterspace} {{\omega _i}}}} \right)^2}\ll1$. In this case, these differential cross sections have the following order of magnitude:
\begin{equation} \label{eq:84}
d{\sigma _0} \sim {Z^2}\alpha r_e^2\left( {\frac{{{\omega _i}}}{m}} \right),\quad d{\sigma _{\left( j \right)res}} \mathbin{\lower.3ex\hbox{$\buildrel<\over
{\smash{\scriptstyle\sim}\vphantom{_x}}$}} \frac{{{Z^2}\alpha r_e^2}}{{\left(\alpha \eta\right)^2 }}{\left( {\frac{{{\omega _i}}}{m}} \right)^3}
\end{equation}

Note that the resonant denominators of the expressions (76),(77) have a characteristic Breit-Wigner form. When $\delta _ + ^2 \to \delta _{ + \left( a \right)}^2$ (for channel a)) and $\delta _ - ^2 \to \delta _{ - \left( b \right)}^2$ (for channel b)) the resonant differential cross sections has a sharp maximum:
\begin{equation} \label{eq:85}
R_{\left( j \right)}^{\max } = \frac{{d\sigma _{\left( j \right)res}^{\max }}}{{d{\sigma _0}}} = {f_0} \cdot {R_{\left( j \right)}},\; {f_0} = \frac{{32{\pi ^3}}}{{{{\left( {\alpha \eta } \right)}^2}}},{\rm{   }}j = a,b
\end{equation}
\begin{equation} \label{eq:86}
R_{(a)} = \frac{{G\left( {{x_ + }} \right)}}{{x_ + ^2{{(1 - {x_ + })}^2}}} \cdot \frac{{{H_{1( + )}}}}{{{H_0} \cdot K_i^2}},
\end{equation}
\begin{equation} \label{eq:87}
R_{(b)}=\frac{{G\left( {{x_ - }} \right)}}{{x_ - ^2{{(1 - {x_ - })}^2}}} \cdot \frac{{{H_{1( - )}}}}{{{H_0} \cdot K_i^2}}.
\end{equation}

Expressions (85) and (86) determine the magnitude of the resonant differential cross section of the PPP (in units of the corresponding differential cross section of the PPP without laser field) for channels a) and b) with simultaneous registration of the radiation angles of the positron and electron (parameters $\delta _ + ^2$  and $\delta _ - ^2$ ), as well as the positron energy in the interval from ${E_{ \left( a \right)+}}$  to $\left[ {{E_{\left( a \right)+}} + d{E_{\left( a \right)+}}} \right]$ (for channel a)) and the electron energy from ${E_{\left( b \right)-}}$ to $\left[ {{E_{\left( b \right)-}} + d{E_{\left(b\right)-}}} \right]$ (for channel b)). It is important to emphasize that for channel a), the positron radiation angle relative to the initial photon momentum (parameter  $\delta _ + ^2$ )  defines both the resonant energy of  positron ${E_{\left( a \right)+}}$ (31) and energy of electron ${E_{\left( a \right)-}} \approx {\omega _i} - {E_{\left( a \right)+}}$. At the same time, these quantities do not depend on the electron radiation angle (parameter $\delta _ - ^2$ ). For channel b) we have the opposite situation. Here radiation angle of the electron relative to the initial photon momentum (parameter $\delta _ - ^2$ ) defines both the resonant energy of  electron ${E_{\left( b \right)-}}$ , and energy of positron ${E_{\left( b \right)+}} \approx {\omega _i} - {E_{\left( b \right)-}}$ . And these quantities do not depend on the positron radiation angle (parameter $\delta _ + ^2$).

From expressions (85), (86) we can see that the magnitude of the maximum resonant cross section, is determined by two functions: $f_0$  , ${R_{\left( a \right)}}$ and ${R_{\left( b \right)}}$. From the one hand the magnitude of the function $f_0$, is mainly determined by the radiation width of the resonance, and is quite large. For the laser wave intensities $\eta  = 0.1$ ($I \sim {10^{16}} \div {10^{17}}\;{{\rm{W}} \mathord{\left/
 {\vphantom {{\rm{W}} {{\rm{c}}{{\rm{m}}^{\rm{2}}}}}} \right.
 \kern-\nulldelimiterspace} {{\rm{c}}{{\rm{m}}^{\rm{2}}}}}$) function ${f_0} \sim {10^9}$. In a real experiment, the width of the resonance will be much larger than the radiation width, and the magnitude of the function $f_0$ will be much less. On the other hand, the magnitude of the functions ${R_{\left( a \right)}}$ and ${R_{\left( b \right)}}$ are determined mainly by the small transmitted momentum when the condition (83) is met and may also be large enough. On Fig. 4.1 and Fig. 4.2 for channels a) and b) represented function ${R_{\left( a \right)}}$ and ${R_{\left( b \right)}}$ dependencies on parameters $\delta _ + ^2$   and  $\delta _ - ^2$  with fixed parameter values  $\delta _ - ^2$ and  $\delta _ + ^2$ correspondently. Solid lines correspond to the energy of a positron (electron) with the sign "+", dotted lines correspond to the energy of a positron (electron) with the sign "-" in front of square root (see (31) and (32)). It can be seen from the figures that when the parameters match $\tilde \delta _ + ^2 = \tilde \delta _ - ^2$  functions ${R_{\left( a \right)}}$ and ${R_{\left( b \right)}}$ have a sharp maximum and can reach 8-11 orders of magnitude. We note that the physical nature of these resonances is determined by the high electron energies and very small transmitted momentum. 
\begin{figure}[h!]
\begin{center}
\begin{minipage}{1.0\linewidth}
     \includegraphics[width=10cm]{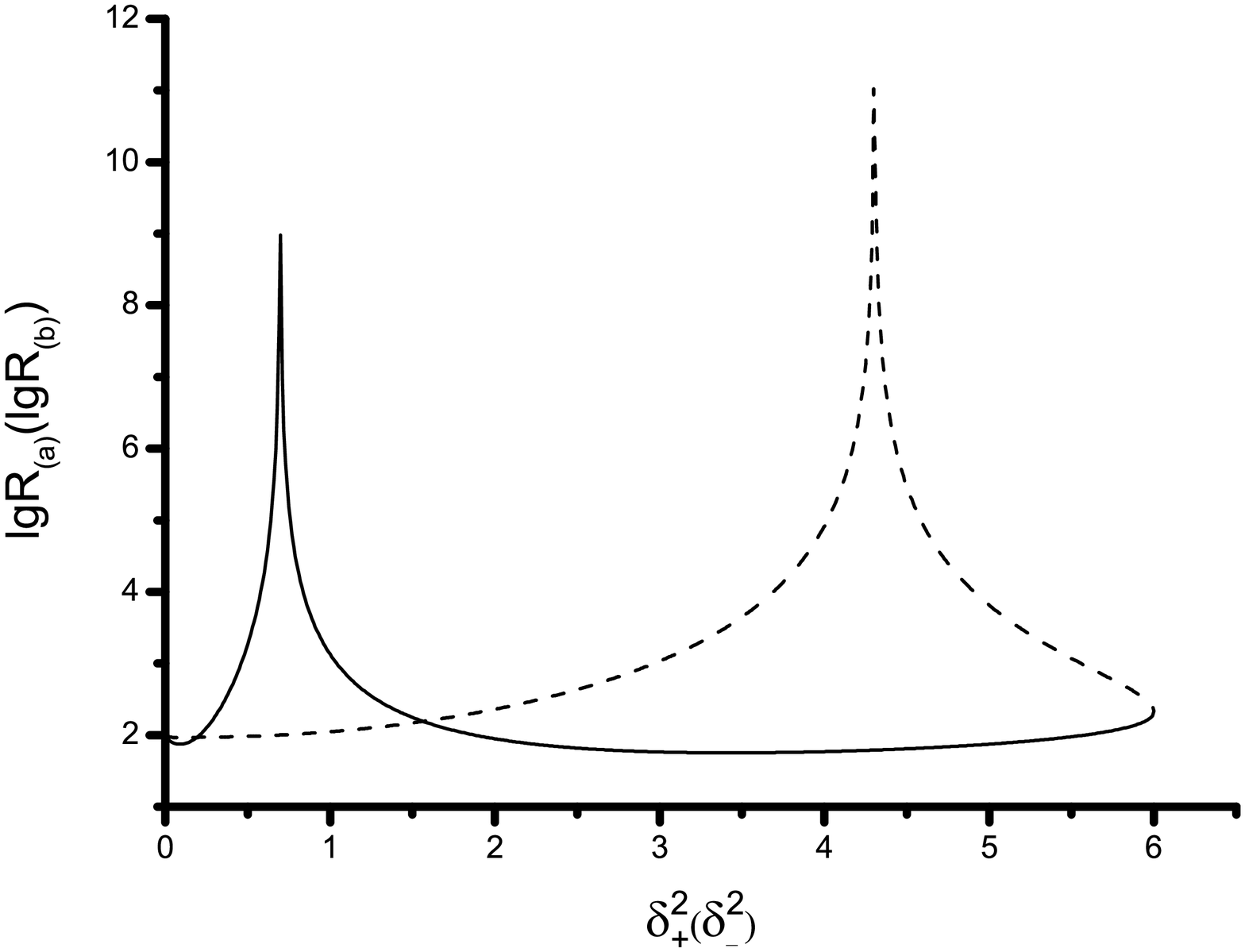}
\caption*{FIG. 4.1: Dependence of the resonant differential cross-section (in units of the corresponding differential cross-section without laser field) on radiation angle plotted for $\tilde \delta _ - ^2 = 1$ (channel a)) and  $\tilde \delta _ + ^2 = 1$ (channel b). Selected parameters: ${\varepsilon _i} = 3,{\rm{ }}{\omega _{thr}} = 83,3{\rm{GeV}}$.}
\end{minipage}
\begin{minipage}{1.0\linewidth}
     \includegraphics[width=10cm]{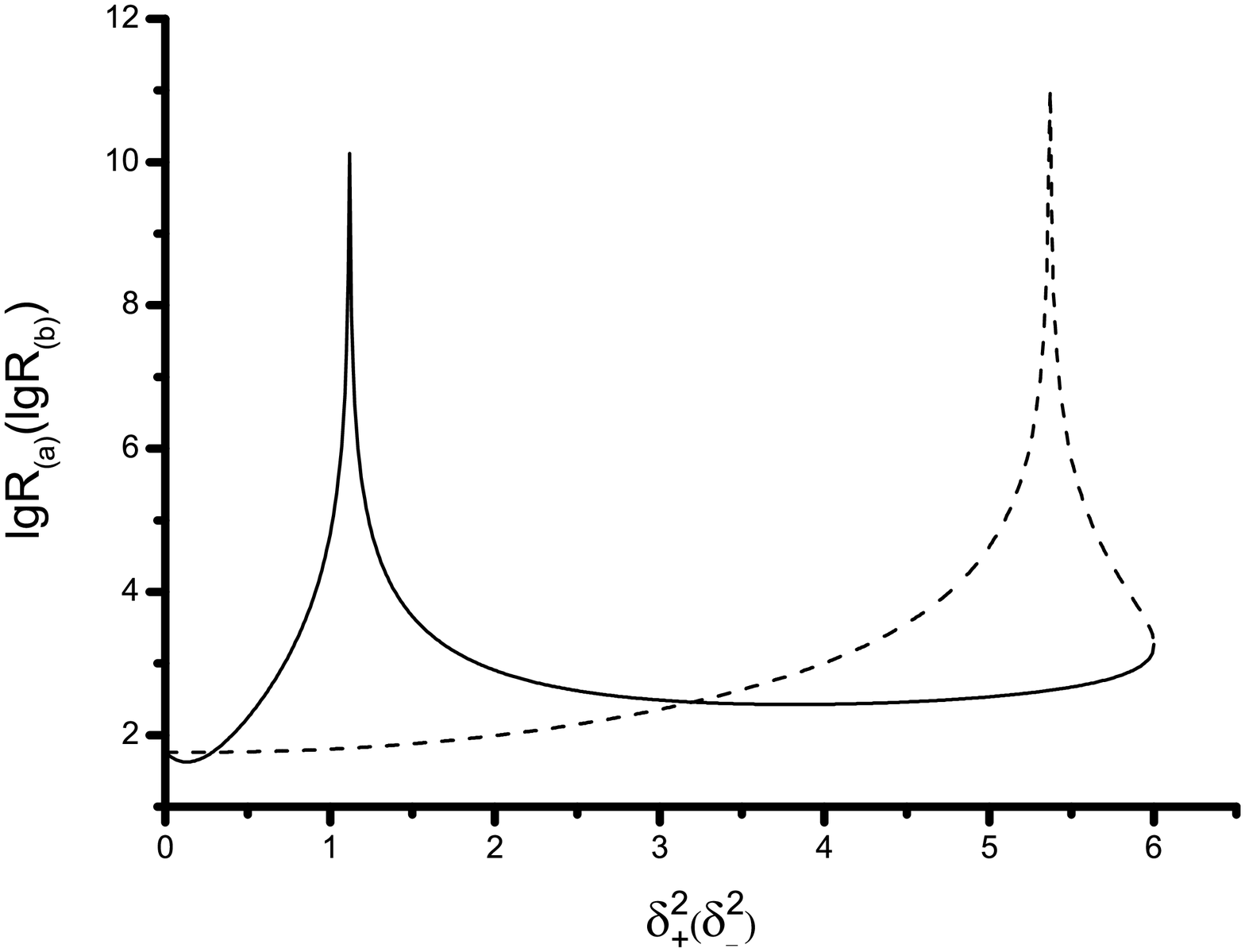}
\caption*{FIG. 4.2:  Dependence of the resonant differential cross-section (in units of the corresponding differential cross-section without laser field) on radiation angle plotted for $\tilde \delta _ - ^2 = 2$  (channel a)) and $\tilde \delta _ + ^2 = 2$  (channel b)). Selected parameters: ${\varepsilon _i} = 3,{\rm{ }}{\omega _{thr}} = 83,3{\rm{GeV}}$.}
\end{minipage}
\end{center}
 \end{figure}

Integrate the resonant differential cross section for channel a) (76) with the respect to $\delta _ - ^2$ (radiation angle of electron) and resonant differential cross section for channel b) (77) with the respect to $\delta _ + ^2$  (radiation angle of positron). After simple calculations we get:
\begin{equation} \label{eq:88}
\begin{aligned}
d\sigma _{(a)}^{res} = \frac{1}{8}\pi \left( {{Z^2}\alpha r_e^2} \right){\left( {\frac{{{\omega _{thr}}}}{m}} \right)^2}\frac{{\left( {1 - {x_{(a) + }}} \right)}}{{{x_{(a) + }}}} \cdot \\
 \cdot \frac{{{\eta ^2}G({x_{(a) + }})}}{{\left[ {{{\left( {\delta _ + ^2 - \delta _{(a) + }^2} \right)}^2} + \Gamma _{{\delta _ + }}^2} \right]}}d{x_{(a) + }}d\delta _ + ^2,
\end{aligned}
\end{equation}
\begin{equation} \label{eq:89}
\begin{aligned}
d\sigma _{(b)}^{res} = \frac{1}{8}\pi \left( {{Z^2}\alpha r_e^2} \right){\left( {\frac{{{\omega _{thr}}}}{m}} \right)^2}\frac{{\left( {1 - {x_{(b) - }}} \right)}}{{{x_{(b) - }}}} \cdot \\
 \cdot \frac{{{\eta ^2}G({x_{(b) - }})}}{{\left[ {{{\left( {\delta _ - ^2 - \delta _{(b) - }^2} \right)}^2} + \Gamma _{{\delta _ - }}^2} \right]}}d{x_{(b) - }}d\delta _ - ^2.
\end{aligned}
\end{equation}

It takes into account that the function ${d_1}\left( {{x_ \pm }} \right) = 4{\varepsilon _i}$ in such kinematics. The resulting expression for the resonant differential cross section (88) determines the angular distribution (and energy (31)) of positron (irrespective to the directions of electron propagation). And the expression for the resonant differential cross section (89) determines the angular distribution (and energy (32)) of electron (irrespective to the directions of positron propagation).

When $\delta _ + ^2 \to \delta _{\left( a \right) + }^2$ (for channel a)) and $\delta _ - ^2 \to \delta _{\left( b \right) - }^2$ (for channel b)) resonant differential cross sections (88) and (89) have sharp maximum and take maximum values.
\begin{equation} \label{eq:90}
d\sigma _{\left( a \right)res}^{\max } = \left( {{Z^2}\alpha r_e^2} \right){g_i} \cdot {F_ + }\left( {{x_{\left( a \right) + }}} \right) \cdot d{x_{\left( a \right) + }} \cdot d\delta _ + ^2,
\end{equation}
\begin{equation} \label{eq:91}
d\sigma _{\left( b \right)res}^{\max } = \left( {{Z^2}\alpha r_e^2} \right){g_i} \cdot {F_ - }\left( {{x_{\left( b \right) - }}} \right) \cdot d{x_{\left( b \right) - }} \cdot d\delta _ - ^2,
\end{equation}
\begin{equation} \label{eq:92}
{g_i} = \frac{{128{\pi ^3}}}{{{{\left( {\alpha \eta } \right)}^2}}}{\left( {\frac{{{\omega _{thr}}}}{m}} \right)^2},
\end{equation}
\begin{equation} \label{eq:93}
\begin{aligned}
{F_ + }\left( {{x_{(a) + }}} \right) = \frac{{{x_{(a) + }}}}{{\left( {1 - {x_{(a) + }}} \right)K_i^2}} \cdot G\left( {{x_{(a) + }}} \right)\\
{F_ - }\left( {{x_{(b) - }}} \right) = \frac{{{x_{(b) - }}}}{{\left( {1 - {x_{(b) - }}} \right)K_i^2}} \cdot G\left( {{x_{(b) - }}} \right)
\end{aligned}
\end{equation}
The resulting expression (90) determines the maximum possible values of the resonant differential cross section for channel a) with simultaneous registration of the radiation angle, as well as the positron energy. At the same time, the expression (91) determines the maximum possible values of the resonant differential cross section for channel b) with simultaneous registration of the radiation angle, as well as the electron energy. 

From the expressions (90),(91) we can see that magnitude of the maximum resonant cross section mainly determined by function $g_i$
(92), which is in optical range of frequencies and for intensities of the laser $\eta  = 0.1$ ($I \sim {10^{16}} \div {10^{17}}\;{{\rm{W}} \mathord{\left/
 {\vphantom {{\rm{W}} {{\rm{c}}{{\rm{m}}^{\rm{2}}}}}} \right.
 \kern-\nulldelimiterspace} {{\rm{c}}{{\rm{m}}^{\rm{2}}}}}$) equals to:
\begin{equation} \label{eq:94}
{g_i} \approx 2,4 \cdot {10^{20}}
\end{equation}

Note that a sufficiently large value of the function $g_i$  is associated not only with the small radiation width of the resonance, which contributes about $ \sim {10^6} \div {10^8}$, but also with very small transmitted momentums in the given resonant process, whose contribution is decisive in $g_i$.

\end{document}